\documentclass[referee]{raa}
\usepackage{graphicx,times}
\usepackage{natbib}
\usepackage{amssymb,amsmath}
\bibpunct{(}{)}{;}{a}{}{,}

\usepackage[a4paper=true,dvipdfm=true,pagebackref=true]{hyperref}
\hypersetup{pdftitle = The title of my PDF, pdfauthor = My name, pdfsubject= The subject, pdfkeywords = keyword1 keyword2 keyword3}
\hypersetup{colorlinks = true, linkcolor = green, anchorcolor = red, citecolor = blue, filecolor = red, pagecolor = red, urlcolor = red}

\begin{document}

   \title{On the Full Spectrum Fitting of Luminous Red Galaxies by Using ULySS and STARLIGHT
$^*$
\footnotetext{\small $*$ Supported by the National Natural Science Foundation of China (NSFC) under grants Nos.11033001 and 11073024.}
}

 \volnopage{ {\bf 2013} Vol.\ {\bf X} No. {\bf XX}, 000--000}
   \setcounter{page}{1}

   \author{Gaochao Liu\inst{1,2,3}, Youjun Lu\inst{1}, Xuelei Chen\inst{1,4}, Wei Du
      \inst{1}, Yongheng Zhao\inst{1}
   }

   \institute{Key Laboratory of Optical Astronomy, National Astronomical Observatories, Chinese Academy of Sciences, Beijing, 100012, China; {\it gcliu@nao.cas.cn}\\
        \and
             University of Chinese Academy of Sciences, Beijing, 100049, China\\
	\and
 College of Science, China Three Gorges University, Yichang 443002, China\\
\and
Center for High Energy Physics, Peking University, Beijing 100871,
China\\
\vs \no
   {\small Received ; accepted }
}
\abstract{In this paper, we select a sample of quiescent luminous red galaxies (LRGs) from Sloan Digital Sky Survey Data Release 7 (SDSS DR7) with signal-to-noise ratio (S/N) substantially high to study the consistency of the full spectrum fitting method by using different packages, mainly, ULySS and STARLIGHT. The spectrum of each galaxy in the sample is fitted by the full spectrum fitting packages ULySS and STARLIGHT, respectively. We find: (1) for spectra with higher S/Ns, the ages of stellar populations obtained from ULySS are slightly older than that from STARLIGHT, and metallicities derived from ULySS are slightly richer than that from STARLIGHT. In general, both packages can give roughly consistent fitting results. (2) for low S/N spectra, it is possible that the fitting by ULySS can be trapped some local minimum parametric regions during execution and thus may give unreliable results, while STARLIGHT can still give reliable results. Based on the fitting results of LRGs, we further analyse their star formation history (SFH) and the relation between their age and velocity dispersion, and find that those highly agrees with conclusions from others' previous work.
}

   \authorrunning{G. C. Liu et al. }            
   \titlerunning{Comparison ULySS with STARLIGHT}  
   \maketitle

%
\section{Introduction}           
\label{sect:intro}
Stellar population synthesis has been widely adopted to study distant galaxies and obtain various galaxy properties. The principle of stellar population synthesis method is to find a combination of series of simple stellar populations (SSPs), of which the theoretical features can match the observational features of a galaxy (\citealt{Crampin+Hoyley+1961}; \citealt{Zhang+2003}). Usually, the best combination of SSPs can be accessible by matching the spectral energy distributions (SEDs), spectral line indices, or full spectrum of the combinations of SSPs with the observed ones.  Fitting via SEDs depends on the shapes of continua and is seriously affected by dust extinction. The utilization of spectral line indices is to make use of strength or equivalent width (EW) of some obvious line features, but it would be difficult to measure the lines which are blended with other lines due to the effect of Doppler line broadening particularly for low-resolution spectra. The full spectrum fitting method takes full advantage of all valuable information contained in a spectrum, including both the continuum and line features. The full spectra fitting method can be execute by different techniques, such as, ULySS and STARLIGHT. It is interesting to check whether the fitting results rely on the choice of the fitting package or not. In this paper, we investigate the differences, if any, between the two full-spectrum fitting packages, i.e., ULySS and STARLIGHT, by fitting spectra of a sample of quiet luminous red galaxies (LRGs). We also use the results to further study the physical properties of LRGs.


\section{Sample Selection}
\label{sect:Sample}

The Sloan Digital Sky Survey (SDSS) is one of the most ambitious and influential surveys in the history of astronomy. It has already completed its two phases of operations (SDSS-I, 2000-2005; SDSS-II, 2005-2008) and obtained deep and multi-color images covering more than a quarter of the sky and meanwhile derived high-quality spectra of a portion of objects. The 3rd phase of operation (SDSS-III, 2008-2014) is currently ongoing . The SDSS uses a dedicated 2.5-meter telescope at Apache Point Observatory, New Mexico, equipped with two powerful special-purpose instruments. The spectrograph used in SDSS-I and SDSS-II are fed by 640 optical fibers of $3^{''}$ entrance aperture and in SDSS-III are fed by 1000 fibers of $2^{''}$ aperture. SDSS releases its data periodically and in this paper we use data from SDSS DR7 released in July, 2008 (\citealt{Abazajian+etal+2009}).

Besides main galaxy sample, LRGs are also the important targets of the SDSS-I and SDSS-II survey. On the one hand, LRGs are supposed to be passively evolving galaxies, so they have simple and pure stellar populations and their spectra can be fitted by simple stellar populations (SSPs). On the other hand, a large sample of LRGs can be obtained as they are bright sources. According to the selection criteria mentioned in \citet{Liu+etal+2012}, we pick out a sample of LRGs which satisfy the following rules to test the two stellar population synthesis packages.

(1) We select galaxies with the ``TARGET-GALAXY-RED" flags which marks those LRGs selected by the algorithm in \citet{Eisenstein+etal+2001} from SDSS database. In this paper, we only choose LRGs that satisfy CUT I (z$<$0.4).

(2) In order to ensure the reliability of the fitting results, we confine the spectroscopic signal-to-noise(S/N) in r-band to be greater than 25. We also select LRGs with spectroscopic S/N$>$10 in r-band in order to check the effects of S/N on the fitting results.

 (3) More strict constraints to our sample. We require the SpecClass EQ 'SPEC-GALAXY' to make sure that the object is a true galaxy , zStat EQ 'XCORR-HIC' to make sure that the spectroscopic redshift of the galaxy is from the cross-identification between spectra and templates, zWarning EQ 0 to make sure that the redshift value is correct, $eClass<0$ to make sure that it is composed with old stellar populations, $z<0.4$ and $fracDew_r>0.8$ to make sure that its surface brightness profile can be well fitted by the de Vaucouleurs curves.

We obtain a sample of 27,695-LRGs from SDSS DR7 that satisfy the above criteria. In order to just use simple stellar populations during the fitting, we need to, above all, ensure that the galaxies in our LRG sample must be those completely evolved. Therefore, we derive fluxes of several spectral lines ($H_{\alpha}$ and $[OII]$) from the published MPA-JHU value-added catalogs \footnotetext{$http://www.map-garching.mpg.de/SDSS/DR7/raw-data.html$} for all galaxies in our sample and then by further selection, we obtain 3,452 spectra with no emissions in $H_{\alpha}$ or $[OII]$ within 2-$\sigma$ level. To further obtain a sample with properties well-distributed, we obtain the velocity dispersions, which are indicators of galaxy mass, for each galaxy of the 3452-galaxy sample from the MPA-JHU value-added catalogs and select galaxies with velocity dispersions between 200 and 320 $km s^{-1}$ to be our final sample. Since there are only 52 galaxies with velocity dispersions greater than 320 $km s^{-1}$ in total, we ingore the 52 galaxies. Hence, our final sample is composed of 2,440 LRGs, which are divided into 4 sub-samples listed in Table 1 in terms of a velocity dispersion interval of 30 $km s^{-1}$.

\begin{table}
\bc
\begin{minipage}[]{150mm}
\caption[]{Quantities of LRGs in each sub-sample of LRGs.\label{tab1}}\end{minipage}
\setlength{\tabcolsep}{1pt}
 \begin{tabular}{l c c c c }
   \hline
Sample& $\sigma_{v}$ & median Redshift &  median Absolute Magnitude &  Number \\
      & ($\rm {km s^{-1}}$)&  & (r band) & \\
 \hline

sub-sample I   & 200 $<\sigma_{v}\le$230 & 0.08 & -20.96 & 791  \\
sub-sample II  & 230$<\sigma_{v}\le$260 & 0.11  & -21.41 & 899  \\
sub-sample III & 260$<\sigma_{v}\le$290 & 0.13  & -21.70 & 553  \\
sub-sample IV  & 290$<\sigma_{v}\le$320 & 0.14  & -21.90 & 197 \\
total          & 200$<\sigma_{v}\le$320 & 0.11  & -21.40 & 2,440 \\
  \hline

  \end{tabular}
\ec
\end{table}

\begin{figure}[!htp]
\centering
\includegraphics [width=10.0cm, angle=0]{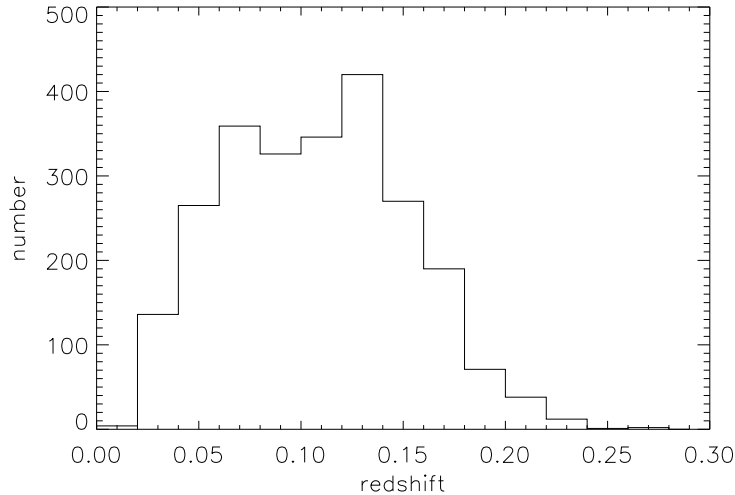}
\caption{
The redshift distribution of the LRGs.}
 \label{fig:f1}
\end{figure}

Figure ~\ref{fig:f1} shows the redshift distribution of our sample and we can see that redshifts of the high S/N spectra (S/N$>$25) in our sample are merely less than 0.25.

\section{Fitting Methods}
\subsection{ULySS}
\begin{figure}[!htp]
\centering
\includegraphics [width=14.0cm, angle=0]{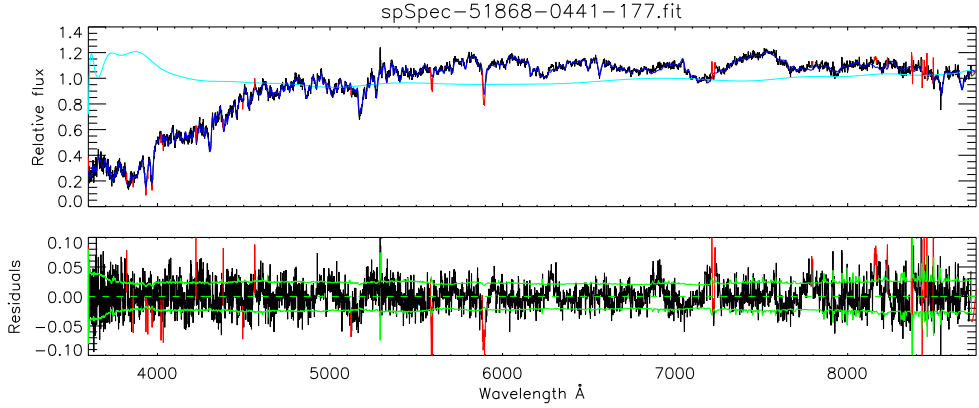}
\caption{The fitting for the spectrum Spec-51868-0441-177.fit by ULySS. In the top panel, the black and blue lines represent the original spectrum and the best fit, respectively, and the cyan line represents the multiplicative n-order polynomial to absorb the effects of the imprecise flux calibration and the Galactic extinction. The pixels in the red region are masked in the fitting. The bottom panel shows the residual spectrum of the best fit and the green lines represent 1-$\sigma$ level.
}
 \label{fig:f2}
\end{figure}

ULySS (\textbf{U}niversity of \textbf{Ly}on \textbf{S}pectroscopic analysis \textbf{S}oftware; \citealt{Koleva+etal+2009}) is an open-source software package written in the GDL/IDL language to analyse astronomical data by a team of Lyon University, France. It can be used for free since the year 2009. ULySS has two powerful functions: (1) it is able to estimate stellar atmospheric parameters (effective temperature-$T_{eff}$, surface gravity-log $g$, metallicity-[Fe/H]) and radial velocity (R$_V$) of stars automatically. (2) it can be used to study stellar populations, star formation histories and chemical enhancement histories of galaxies and star clusters. As a stellar population synthesis tool, ULySS,  fits the full spectrum of a stellar system with a linear combination of multiple simple stellar populations (SSPs). By minimizing $\chi^2$ of the fitting, it resolves the most probable SSPs and relevant physical properties of the galaxy. An observational spectrum can be expressed as a linear weighted ($W$) combination of $k$ non-linear components (CMP) convolved with a line-of-sight velocity distribution (LOSVD) and multiplied by a polynomial continuum ($n$-order polynomials) and added by another polynomial ($Q_m(\lambda)$), i.e,

\begin{align}
 F_{obs}(\lambda) = P_{n}(\lambda) \times \bigg(&{\rm LOSVD}(v_{sys},\sigma, h3, h4) \nonumber \\
    &\otimes \sum_{i=0}^{i=k} W_i \,\, {\rm CMP}_i\,(a_1, a_2, ...,\lambda)\bigg) + Q_{m}(\lambda),
\label{eq:main}
\end{align}
where LOSVD is a function of the systematic velocity,$v_{sys}$, the velocity dispersion, $\sigma$ and perhaps include Gauss-Hermit expansion ($h3$ and $h4$), $\lambda$ here is the logarithm of the wavelength, The CMP$_i$ has distinguished expressions for different problems. For example, if ULySS is used to study stellar atmospheric parameters, CMP$_i$ will be a function of $t_{eff}$, log $g$, and [Fe/H]. If ULySS is used to study stellar populations of galaxies or star clusters, CMP$_i$ will be a function of age, [Fe/H] and [Mg/Fe]. The n-order polynomial, $P_{n}(\lambda)$, absorbs the effects of imprecise flux calibration and the Galactic extinction,
which is determined by ordinary least-squares at each evaluation of
the function minimized by the Levenberg-Marquardt routine and we can not decide
the shape of $P_{n}(\lambda)$ before the spectra fitting.
The additive polynomial, $Q_{m}(\lambda)$, is certainly more subtle to use, and is, in most cases, unnecessary. The ULySS package uses the full spectrum instead of only some characteristic spectral lines to avoid the errors due to measurements of only individual lines and also avoid the degeneracies between the fitting results of some physical parameters by estimating the required parameters simultaneously. ULySS also introduces a Line-Spread-Function (LSF) and matches the resolution of the observational spectrum and model spectrum by injecting the relative LSF  into either the model spectrum or observational spectrum. A spectrum (spSpec-51868-0441-177.fits) from our sample as an example in Figure ~\ref{fig:f2} to show the fitting by ULySS.

\subsection{STARLIGTH}

\begin{figure}[!htp]
\centering
\includegraphics[width=14.0cm, angle=0]{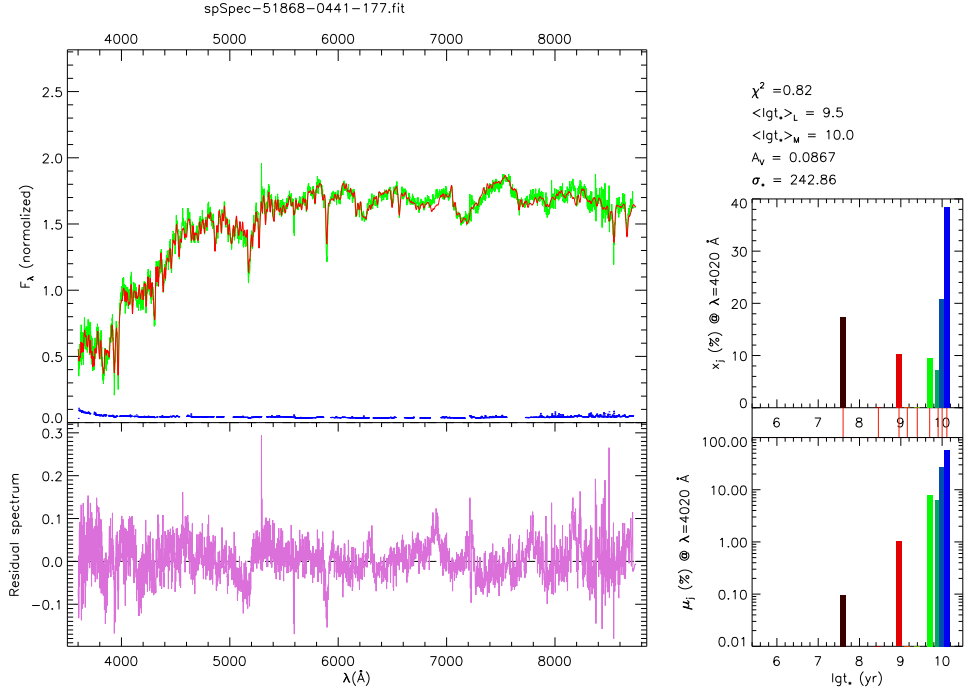}
\caption{The fitting for the spectrum spSpec-51868-0441-177.fits by STARLIGHT. In the upper left, the green stands for the original spectrum, the red for the best-fit spectrum, and the blue for the errors in which the missing place is not used to fit. The bottom left panel shows the residual spectrum of the best fit. The plots in the upper right and the right bottom respectively exhibit the light fraction ($x_j$) and mass fraction ($\mu_j$)of individual stellar populations. The middle grid between the two plots in the right shows ages of the SSPs.
}
 \label{fig:f3}
\end{figure}

STARLIGHT (\citealt{Cid+etal+2005}) is one of the most powerful tools for stellar population studies, which is widely-used in the field of studies on early-type galaxies, late-type galaxies, star clusters and AGNs. STARLIGHT fits an observational spectrum with a model spectrum, $M_{\lambda}$, which is a combination of $N_\star$ SSPs with different ages and metallicites defined by users subjectively.  $M_\lambda$ can be expressed as follows

\begin{equation} \label{Mlambda} M_\lambda = M_\lambda(\vec{x}, A_V, v_\star, \sigma_\star) =
\sum_{j=1}^{N_\star} x_j \gamma_{j,\lambda} r_\lambda,
\end{equation}
where, $\gamma_{j,\lambda} \equiv b_{\lambda,j} \otimes G(v_\star, \sigma_\star)$, $b_{\lambda,j} \equiv
{B_{\lambda,j}\overwithdelims () B_{\lambda_{0},j}}$ refers to the normalized flux of the jth observational spectrum,
$B_{\lambda,j}$ refers to the flux of the jth model spectrum, $B_{\lambda_{0},j}$ is the flux of the jth normalized model spectrum at the wavelength of $\lambda_0$, at which the flux has been used to normalize the full model spectrum, $G(v_\star, \sigma_\star)$ describes the stellar movement at the radial direction and is expressed by a Gaussian function with a center at $v_\star$ and a width of $\sigma_\star$.
$r_\lambda \equiv 10^{-0.4(A_\lambda-A_V)}$ represents the global extinction. The $x_j (j = 1; \cdots;N)$, as one of the most significant parameters among all STARLIGHT outputs, stands for contribution fraction of the SSP component with equivalent age of $t_j$ and metallicity of $Z_j$ to the flux of model spectrum at the normalization wavelength $\lambda_0$=4020$\rm \AA$. Similarly to $x_j$, $\mu_j$ stands for the contribution fraction to stellar mass.
The core algorithms of STARLIGHT includes the Simulated Annealing Algorithm and the Metropolis Algorithm of Markov Chain-Monte Carlo Method, which search the optimized parametric results in the whole parameter space to minimize $\chi^2$ value. The $\chi^2$ can be expressed as $\chi^2 = \sum_\lambda [(O_\lambda- M_\lambda) w_\lambda]^2$, where $w_\lambda^{-1}$ stands for errors of the observational spectrum. According to formulas of

\begin{equation}
\label{logtl}
\langle {\rm log} t_\star \rangle_L = \sum _{j=1}^{N_\star} x_j \, {\rm log} \, t_j,
\end{equation}

\begin{equation}
\label{logtm}
\langle {\rm log} t_\star \rangle_M = \sum _{j=1}^{N_\star} \mu_j \, {\rm log} \, t_j,
\end{equation}

and

 \begin{equation}
  \label{zl}
  \langle Z \rangle_L = \sum _{j=1}^{N_\star} x_j z_j
  \end{equation}
We can obtain light-weighted age and metallicity of the galaxy.
Figure ~\ref{fig:f3} shows the STARLIGHT fitting to the same spectrum as that shown in Figure ~\ref{fig:f2}.

\subsection{Template library}

There are a variety of template spectral libraries for evolutionary stellar population synthesis, such as BC03, Ma05, GALEV, GRASIL, Vazdekis/Miles and so on. \citet{Chen+etal+2010} used STARLIGHT to investigate the effect
 due to the choice of different template libraries on the stellar population synthesis and found that there are still some differences in the output between these different template libraries. In this paper, in order to compare the differences between the two individual packages, we use the widely-used template library, BC03 (\citealt{Bruzual+Charlot+2003}), fixed in the whole fitting work. The BC03 model has 6900 points in wavelength from $91\AA\sim 160 \mu{}m$, 221 points in age from 0$\sim$20 Gyr, and 6 points in metallicity of 0.0001,\quad 0.0004,\quad 0.004,\quad 0.008,\quad 0.02,\quad 0.05. A SSP spectrum in BC03 is generated from the stellar spectral library, SteLib, and has a resolution of FWHM=3 $\AA$ in the optical wavelength range from 3200 to 9500 $\AA$.
In our work, we choose Chabrier Initial Mass Function (IMF) (\citealt{Chabrier+2003}) and Padova94    (\citealt{Bertelli+etal+1994}) stellar evolutionary tracks.

We list main parameters of BC03 in Table 2.

\begin{table}
\bc
\begin{minipage}[]{150mm}
\caption[]{Main parameters of BC03 model.\label{tab2}}\end{minipage}
\setlength{\tabcolsep}{1pt}
 \begin{tabular}{l c c c c c c c }
  \hline
Model & Stellar library& Resolution & Wavelength  & Age  & Metallicity  &  IMF& Stellar evolutionary track \\
& &(\AA) & (\AA) & (Gyr) & (dex) & &  \\
 \hline
 BC03 & SteLib & 3 & 3200-9500 & 0.1-20 & $-2.3$-0.4 & Chabrier & Padova~94\\
  \hline

  \end{tabular}
\ec
\end{table}

Using STARLIGHT, users need to choose SSPs with required ages and metallicities from the model.
Since LRGs are regarded as old-aged and metal-rich galaxies, we choose 9 values of age (4,\ 286,\ 900~Myr£¬\ 1.4,\ 2.5,\ 5,\ 8,\ 10,\ 13~Gyr) and 3 values of metallicity (0.004,\ 0.02,\ 0.05) to construct 27-SSP model spectra so as to decrease the uncertainty of fitting results which would increase with the numbers of SSPs we use. Only one need be paid attention is that
the choice of age and metallicity values of base spectrum must cover all possibility of parameter space of study samples, at the same time, the number of base spectrum as small as
possible to reduce the uncertainty of fitting results. We also choose 10 values of age (i.e.
4,\ 286,\ 900~Myr£¬\ 1.4,\ 2.5,\ 5, \ 7, \ 9,\ 11,\ 13~Gyr) and 4 values of metallicity (i.e.
 0.004, 0.008, 0.02, 0.05) to construct 40-SSP model spectra to fit the spectra of our sample, but the fitting results only change slightly.
The direct fitting results of STARLIGHT include the contribution to light ($x_j$) and mass ($\mu_j$) of every SSP and extinction. While in ULySS package, it assembles all SSP spectra of BC03 into a fit-format file of 5979$\times$116$\times$6 dimension, where 5979, 116, and 6 respectively refers to the number of flux pixels in every SSP spectrum, the number of age points and the number of metallicity points. Other points can be obtained by interpolation between the grid points. Once fitting, STARLIGHT and ULySS all will automatically search the closest model spectrum to the observational spectrum and give the fitting results.

\section{Comparison of results}

The spectrum of each LRG in our sample has been fitted  by STARLIGHT and ULySS, respectively. In this section, we compare the results obtained from ULySS with that from STARLIGHT and check whether the two methods give consistent result or not.

\subsection{Age distribution}

\begin{figure}[tbph]
 \centering
 \includegraphics[width=14.0cm, angle=0] {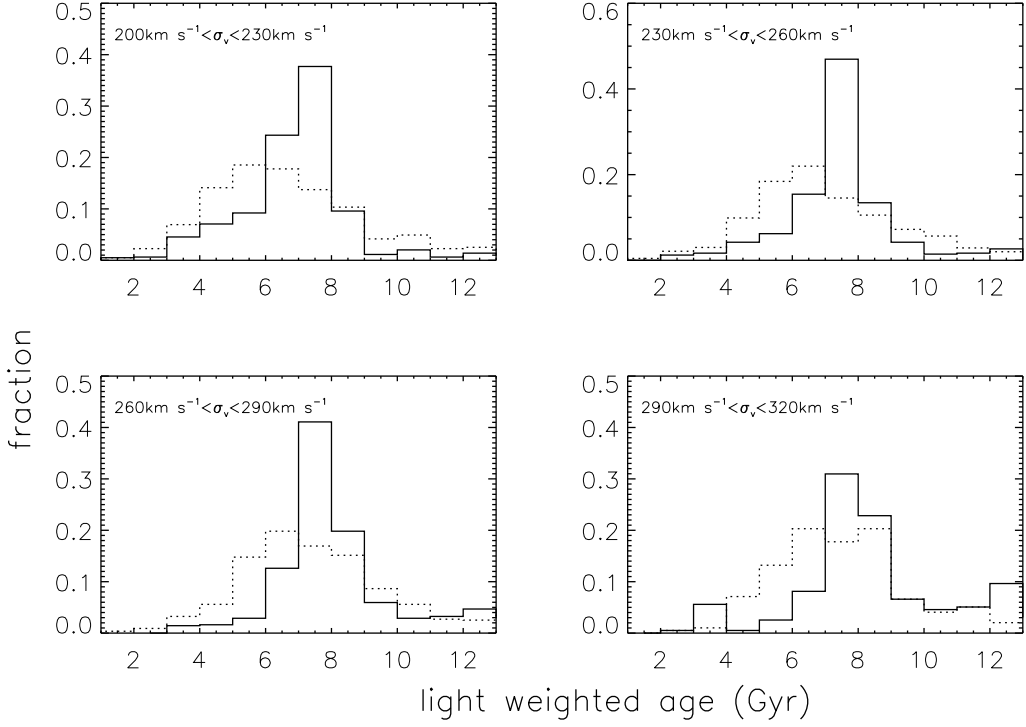}

   \caption{Distributions of the light-weighted ages of LRGs. Solid and dashed lines represent results obtained from ULySS and STARLIGHT, respectively.}
 \label{fig:f4}
 \end{figure}

Since ULySS can only give light-weighted ages for resultant SSPs, we just take the light-weighted ages from STARLIGHT results, although STARLIGHT can additionally give the mass-weighted ages for resultant SSPs.
Figure ~\ref{fig:f4} shows the distributions of the SSPs' light-weighted ages derived from ULySS (solid lines) and STARLIGHT (dashed lines) for the 4 sub-samples, respectively.

 For any of the 4 sub-samples shown in Figure ~\ref{fig:f4}, the light-weighted ages of LRGs derived from STARLIGHT are slightly younger than that derived from ULySS. In addition, both results from STARLIGHT and ULySS suggest that the LRGs sub-sample with larger velocity dispersion ($\sigma$) has larger ages. This conclusion agrees with the 'downsizing' formation of the galaxies, which means that the more massive a galaxy is and the earlier it would have started to form. We tabulate the means and standard deviations of the light-weighted ages derived from STARLIGHT and ULySS for the 4 sub-samples of LRGs.

\begin{table}
\bc
\begin{minipage}[]{150mm}
\caption[]{Mean light-weighted ages of LRGs \label{tab3}}\end{minipage}
\setlength{\tabcolsep}{1pt}
 \begin{tabular}{l c c }
   \hline
Sub-sample & Mean age from STARLIGHT& Mean age from ULySS\\
             & (Gyr) & (Gyr)\\
  \hline

Sub-sample I   & 6.7$\pm$2.4 & 7.0$\pm$1.9 \\
Sub-sample II  & 7.1$\pm$2.3 & 7.6$\pm$1.8 \\
Sub-sample III & 7.6$\pm$2.3 & 8.3$\pm$2.0 \\
Sub-sample IV  & 7.6$\pm$2.1 & 8.6$\pm$2.4\\
  \hline

  \end{tabular}
\ec
\end{table}

\subsection{Metallicity distribution}

\begin{figure}[!htp]
\centering
\includegraphics[width=14.0cm, angle=0] {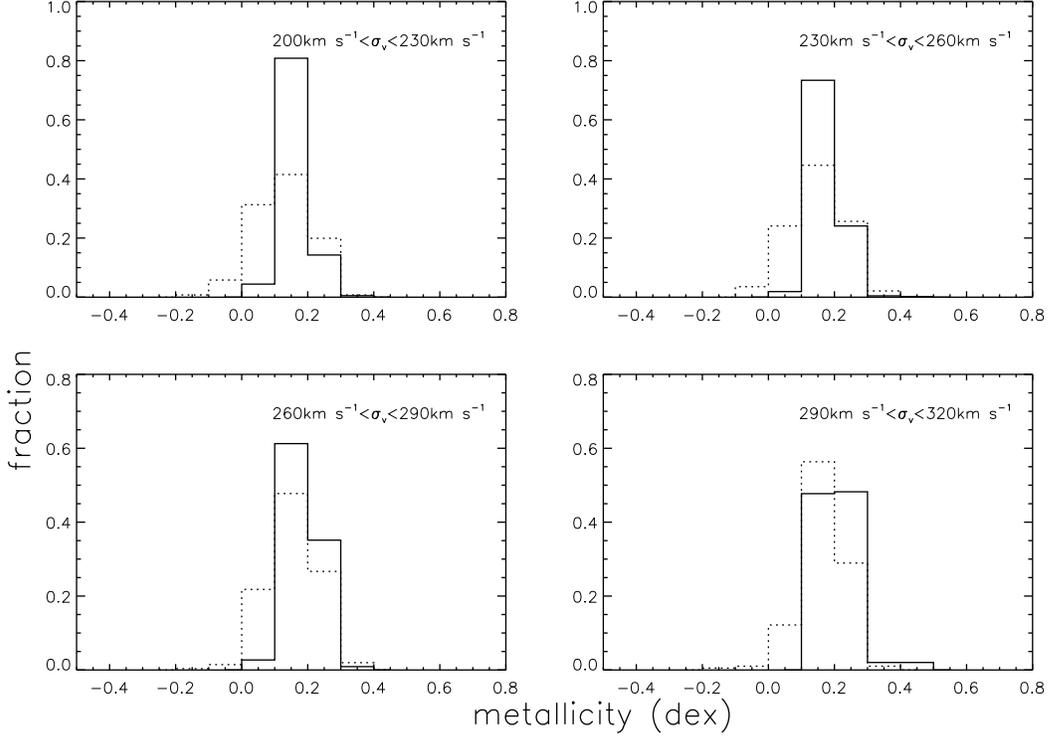}

\caption{Metallicity distribution of LRGs for the four sub-samples, respectively. The solid and dashed lines represent the results obtained from ULySS and STARLIGHT, respectively.}
\label{fig:f5}
\end{figure}

We compare distributions of LRG metallicites derived from STARLIGHT and ULySS in Figure ~\ref{fig:f5}. We can conclude that the majority of LRGs have richer metallicities than the sun, which is in accord with our previous knowledge of LRGs being metal-rich galaxies. Although ULySS gives a bit higher metallicities than STARLIGHT, results from the two packages are generally consistent with each other. We tabulate the means and standard deviations of the light-weighted metallicities derived from STARLIGHT and ULySS for the 4 sub-samples of LRGs.

\begin{table}
\bc
\begin{minipage}[]{150mm}
\caption[]{Mean metallicities of LRGs \label{tab4}}\end{minipage}
\setlength{\tabcolsep}{1pt}
 \begin{tabular}{l c c }
   \hline
Sub-sampe & Mean metallicity from STARLIGHT& Mean metallicity from ULySS\\
             & (dex) & (dex)\\
   \hline

Sub-sample I   & 0.12$\pm$0.09 & 0.16$\pm$0.04 \\
 Sub-sample II  & 0.15$\pm$0.08 & 0.17$\pm$0.04 \\
 Sub-sample III & 0.16$\pm$0.08 & 0.18$\pm$0.04\\
 Sub-sample IV  & 0.16$\pm$0.07 & 0.20$\pm$0.05\\
   \hline

  \end{tabular}
\ec
\end{table}

\subsection{Effects of S/N on fitting results}

\begin{figure}[!htp]
 \centering
 \includegraphics[width=14.0cm, angle=0] {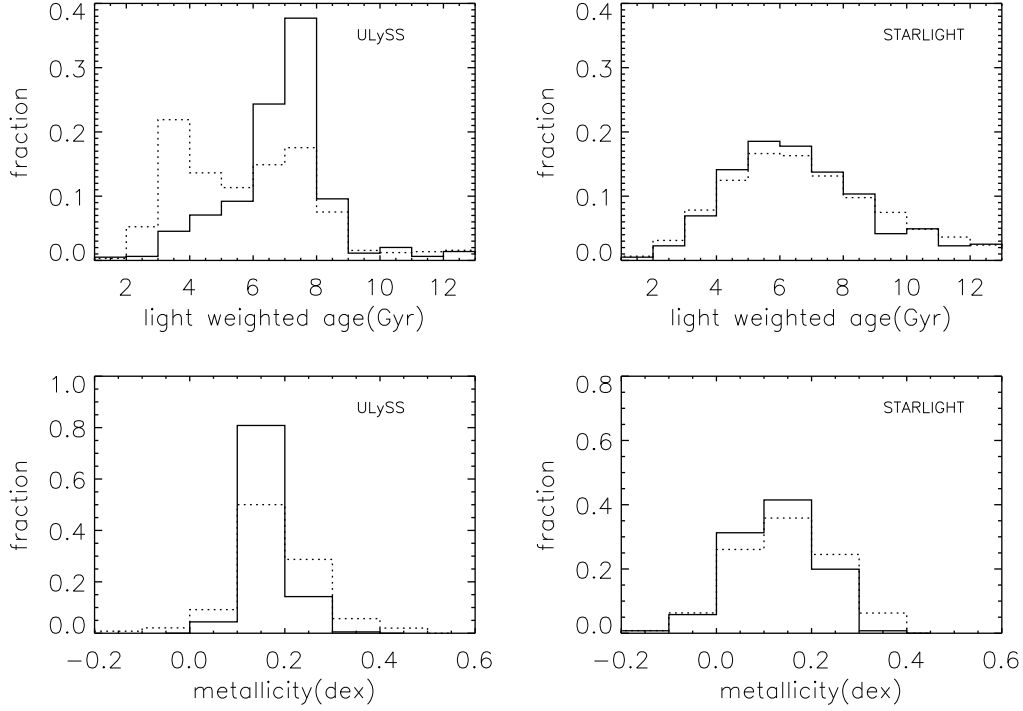}
\caption { Effects of S/N on the fitting results. Top panels show the age distribution obtained by using ULySS (left) and STARLIGHT (right) to fit those LRGs with S/N$>$25 (solid lines) and S/N$>$10 (dashed lines), respectively. The bottom panels show the metallicity distributions.}
\label{fig:f6}
    \end{figure}

In the analysis above, S/Ns of our sample are greater than 25. In order to check the effect of S/N on fitting results from the two packages, we pick out a LRG sample with S/N$>$10 by using the same selection criteria as that in Section 2 but lower the spectral S/N limit down to 10. Then, similarly to what we have done in Section 2, this sample is divided into 4 sub-samples in terms of velocity dispersion, which, respectively, have 4756, 8748, 7149, and 3230 LRGs.
We only take results of the first sub-sample as an example to demonstrate whether spectral S/Ns have influences on fitting results, as conclusions from check on other 3 sub-samples are completely the same as that from the first sub-sample. We show the ages from ULySS fitting for the first sub-sample previously with S/N$>$25(Sample-25; solid line) and currently with S/N$>$10 (Sample-10; dashed line) in the upper left plot in Figure ~\ref{fig:f6}. From this plot, we discover that Sample-10 has two age peaks at nearly 3 Gyr and 7 Gyr. We carefully check fitting processes for spectra having 3 Gyr peaks by performing 500 Monte-Carlo simulations and finally find that the ULySS algorithms for these spectra have all trapped into local minimum regions of the parameter space. Because of easily trapping into local minimum regions of parameter space, ULySS provides additional tools of Monte-Carlo simulations, $\chi^2$ maps, and covergence maps to explore parameter space and check reliabilities of its fitting solutions. Combining our test and its characteristics, ULySS may be more appropriate to spectra with high S/N, but if ULySS is used for spectra with low S/N, the results must be carefully checked by the provided tools (Monte-Carlo simulations, $\chi^2$ maps, and covergence maps) to see whether the fitting has trapped into local minimum regions of parameter space. Similarly, we show the results from STARLIGHT fitting in the upper right plot in Figure ~\ref{fig:f6}, which indicates a consistent conclusion with that from ULySS. We show the effect of S/N on metallicities derived from the two packages in the bottom panel in Figure ~\ref{fig:f6}, from which we conclude that spectral S/N has little effect on metallicity from the fitting.

\section{Validation of fitting methods}
According to the above comparisons, we find that ULySS and STARLIGHT give quantatively
different fitting results, though the general trends of the obtained age distribution and
 metalicity distribution are roughly consistent with each other for those
 spectra with high S/Ns. Which one is better?
To answer this question, we carry out Monte-Carlo simulations with a series know inputs,
and then compare the outputs from two fitting codes with these inputs to address which code
yields more reliable results than the other.

To do this, we select a series model spectra from BC03 template library,
such as, the ages of model spectra are from 1 to 13 ~Gyr with a step 1 ~Gyr
and the metallicities are 0.0001, 0.0004, 0.004, 0.008, 0.02, 0.05, separately.
We select 13$\times$6 model spectra as our test sample totally.
We allocate the model spectra an error spectra whose amplitude equal to the
flux of model spectra divided by S/N and follow the Gaussian distribution. Here,
we assume S/N=25. Next, we fit these model spectra
with ULySS and STARLIGHT, respectively. The fitting is made starting from 3900 $\AA$ to 8500 $\AA$,
which is similar to the wavelength coverage of our LRGs sample.

\begin{figure}[!htp]
 \centering
 \includegraphics[width=16.0cm, angle=0] {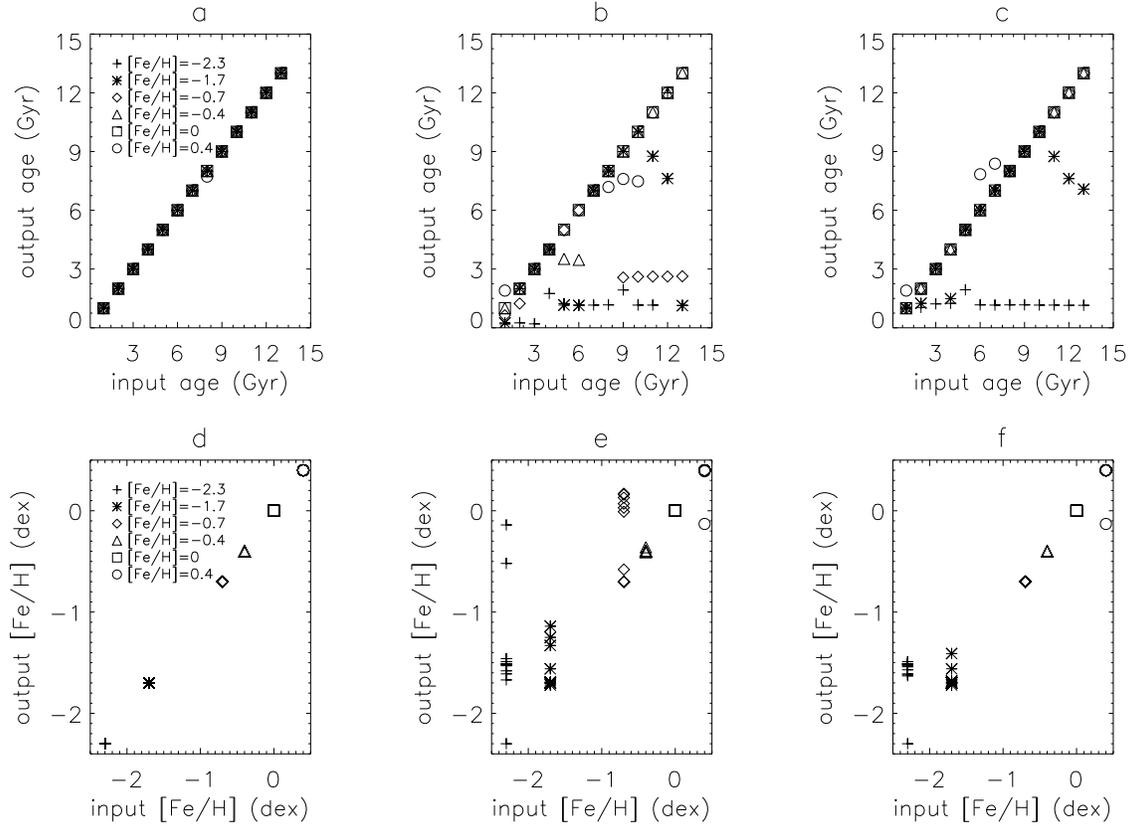}
   \caption{Dependence of the fitting results on the initial estimated value for ULySS.
   Panels (a), (b) and (c) show the fitting age of the model spectrum when IEV of metallicities
   are equal to those inputs of the model spectra, IEV of metallicity is equal
   to 0 dex and IEV of metallicity is equal to -0.7 dex, respectively.
   Panels (d), (e), (f) show the fitting metallicity of the model spectrum correspondingly.
   X-axis represent the inputs and y-axis represent the outputs.
   The plus symbols, asterisks, diamonds, triangles, squares and circles
   represent the fitting results of model spectra with six different
   metallicities from low to high, separately.
     }
\label{fig:f7}
\end{figure}

ULySS uses Levenberg$¨C$ Marquardt routine to search the parameter space to
get the minimization of $\chi^2$, which needs some initial
estimated value (IEV) to begin searching the parameter space. We find the
IEV is crutial to the final fitting results.
Figure ~\ref{fig:f7} shows the fitting results depending on the IEV.
Panels (a) and (d) show the ages and metallicities derived from ULySS fitting
when IEV of metallicities are equal to those input of the model spectra.
We find the outputs almost the same as the inputs.
Panels (b) and (e) show the outputs when IEV of metallicities are equal to 0 dex.
We find the outputs of those spectra with low metallicity (i.e. -2.3,-1.7,-0.7)
far away from the inputs, but the outputs of those spectra around solar metallicity
are well consistent with the inputs.
We also test the situation when IEV of metallicities are equal to -0.7 dex and panel
(c) and (f) shows the results, which are some similar with Panels (b) and (e).
We also find the age and metallicity derived from fitting are anti-correlated by
comparing the outputs of age with that of metallicity, that is,
if we overestimate the metallicity then the age will be underestimated.

\begin{figure}[!htp]
 \centering
 \includegraphics[width=16.0cm, angle=0] {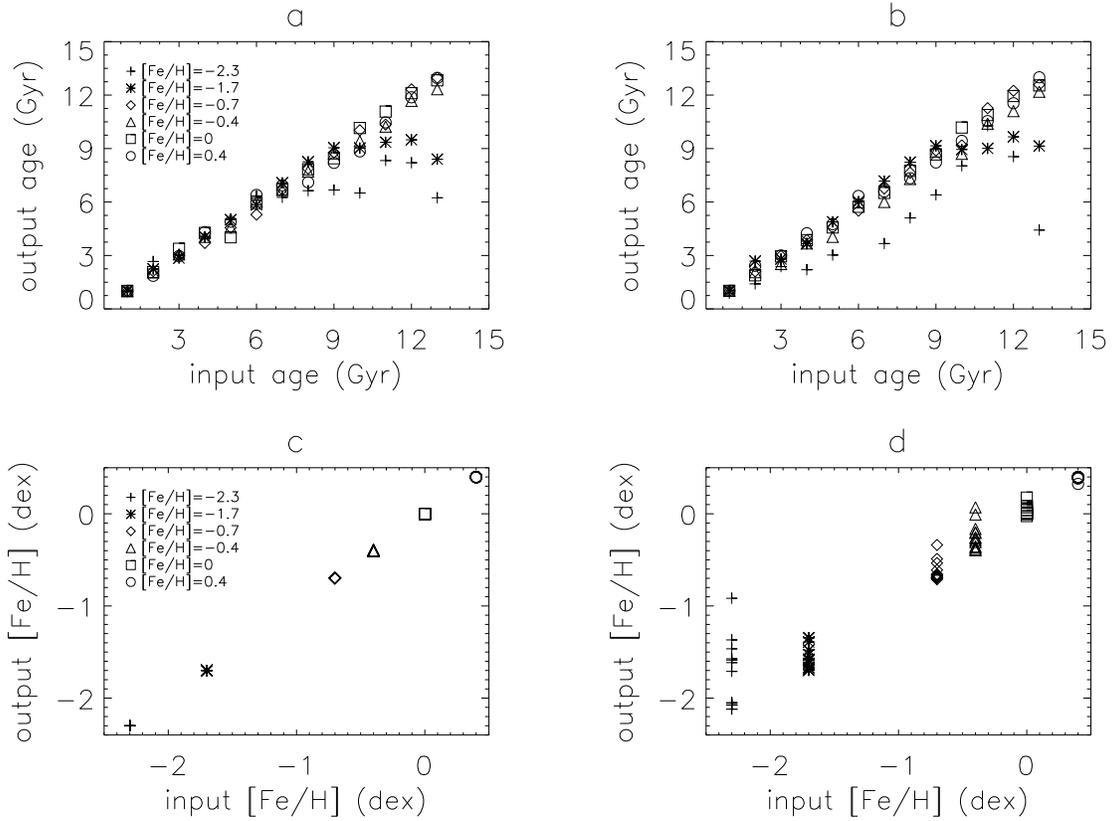}
   \caption{ Fitting results with STARLIGHT.
    Panels (a) and (c) show the fitting results of age and
     metallicity with 8-ssp model spectra, respectively.
 While panels (b) and (d) show the results with 48-ssp model spectra.
   X-axis represent the inputs and y-axis represent the outputs.
   The plus symbols, asterisks, diamonds, triangles, squares and circles
   represent the fitting results of model spectra with six different
   metallicities from low to high, separately.
     }
\label{fig:f8}
\end{figure}

For STARLIGHT, we need to predefine the base spectra. We choose 8 values of age
(500~Myr£¬\ 1,\ 3,\ 5,\ 7,\ 9,\ 11,\ 13~Gyr) and 6 values of metallicity
(0.0001,\ 0.0004,\ 0.004,\ 0.008,\ 0.02,\ 0.05) to construct 48-SSP model spectra.
We also choose 8 values of age and 1 value of metallicity to fit the model spectra
with the same metallicity as our chose value.
Figure ~\ref{fig:f8} shows the fitting results. Panels (a) and (c) show the
fitting results of age and metallicity with 8-ssp model spectra, respectively.
Correspondingly, panels (b) and (d) show the results with 48-ssp model spectra.
From Figure ~\ref{fig:f8}, we can find the fitting results are stable
and robust by adopting either 8-ssp model spectra or 48-ssp model spectra,
though the outputs of age systemic slightly younger than the inputs.
The spectra with very low metallicity also can not be well fitted, which maybe
caused by the limitation of BC03 library since the metallicity bin, [Fe/H]=-1.7 dex
and [Fe/H]=-2.3 is computed from very small stars while spanning a large range in metallicity.

In a word, ULySS and STARLIGHT all can give consistent outputs compared with inputs for those
spectra around the solar metallicity and the fitting results derived from STARLIGHT are more
stable and robust than that from ULySS, since the fitting results derived from ULySS show some
dependence on the initial estimated value.

\section{Properites of LRGs}

By analyzing a sample of LRGs simultaneously with ULySS and STARLIGHT, we make a conclusion that the two packages are able to give generally consistent fitting results, although the ages and metallicities derived from ULySS seem to be a little older and richer than those derived from STARLIGHT. It is worth being noticed that the results should be carefully checked to avoid being trapped into some local minimum regions of parameter space when ULySS is used for low S/N spectra. Basing on parameters derived above, we will further study some physical properties of LRGs in this section.

\subsection{Age-$\sigma$ relation of LRGs}

\begin{figure}[!htp]
 \centering
 \includegraphics[width=10.0cm, angle=0] {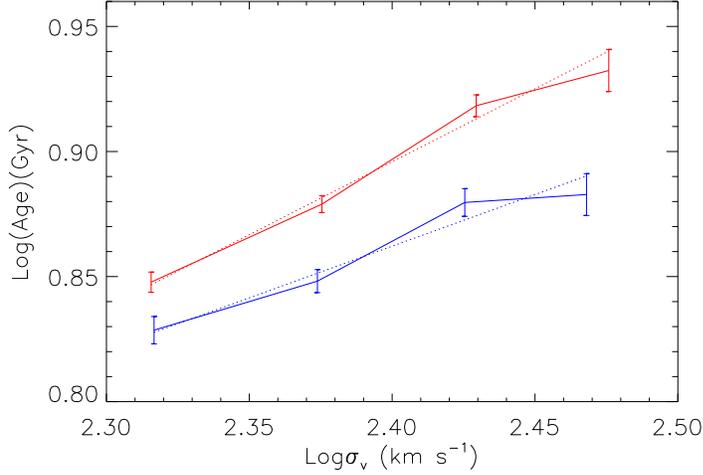}

   \caption{ Age-$\sigma$ relation of LRGs. The red and blue solid lines show the relations obtained from ULySS and STARLIGHT, respectively. The dotted line show the best fit to the Age-$\sigma$ relation of LRGs. }
\label{fig:f9}
\end{figure}

Figure ~\ref{fig:f4} shows that the ages of LRGs increase with increasing velocity dispersions. We give out a quantitative relation between age and $\sigma$ below. Assuming that the increases of age with increasing $\sigma$ follows power low, i.e., $t_{age} \propto \sigma_v^{\gamma}$, then, we average the ages and metallicities for each of the 4 sub-samples, and estimate $\gamma$ by fitting the exponential relation using the mean ages and mean $\sigma$s from the 4 sub-samples. The final age-$\sigma$ relations from ULySS fitting (red line; $\gamma$=0.58$\pm$0.04) and STARLIGHT fitting (blue line; $\gamma$=0.41$\pm$0.06) are shown in Figure ~\ref{fig:f9}. In the above fitting, however, the obtained ages of different galaxies may
 have different reference point since those galxies are
located at different redshifts. Here we correct this by converting the age of each galaxy at
redshift z to that at redshift 0 by adding look back time of the galaxy at z. To do so,
a flat $\Lambda CDM$ cosmology is adopted, i.e., $H_0=70 \rm km~s^{-1}~Mpc^{-1}$ and $\Omega_m=0.30$.
With these corrected ages, we refit the age-$\sigma$ relation, and find
$\gamma$=0.69$\pm$0.04 and $\gamma$=0.56$\pm$0.05 for the relation obtained
from the ULySS and STARLIGHT fitting, respectively.

Our results clearly show the dependence of age on $\sigma$ for early-type galaxies,
as also revealed by a number of other studies. For example, \citet{Nelan+etal+2005} studied the age-$\sigma$ relation of 4097 red-sequence galaxies in 93 low-redshift galaxy clusters, and derived a relation of  $t_{age}\propto \sigma_v^{0.59\pm0.13}$; \citet{Simth+etal+2009} investigated a sample of 232 quiet galaxies in Shapley super cluster of galaxies, and obtained a relation of $t_{age}\propto \sigma_v^{0.40}$. However, the estimates of $\gamma$ seem to be slightly different in different studies, which might be due to different sample selection criteria, different treatment for emission lines, and different fitting methods etc.( as discussed in \citet{Nelan+etal+2005})

\subsection{Star formation history of LRGs}
\begin{figure}[!htp]
 \centering
 \includegraphics[width=14.0cm, angle=0] {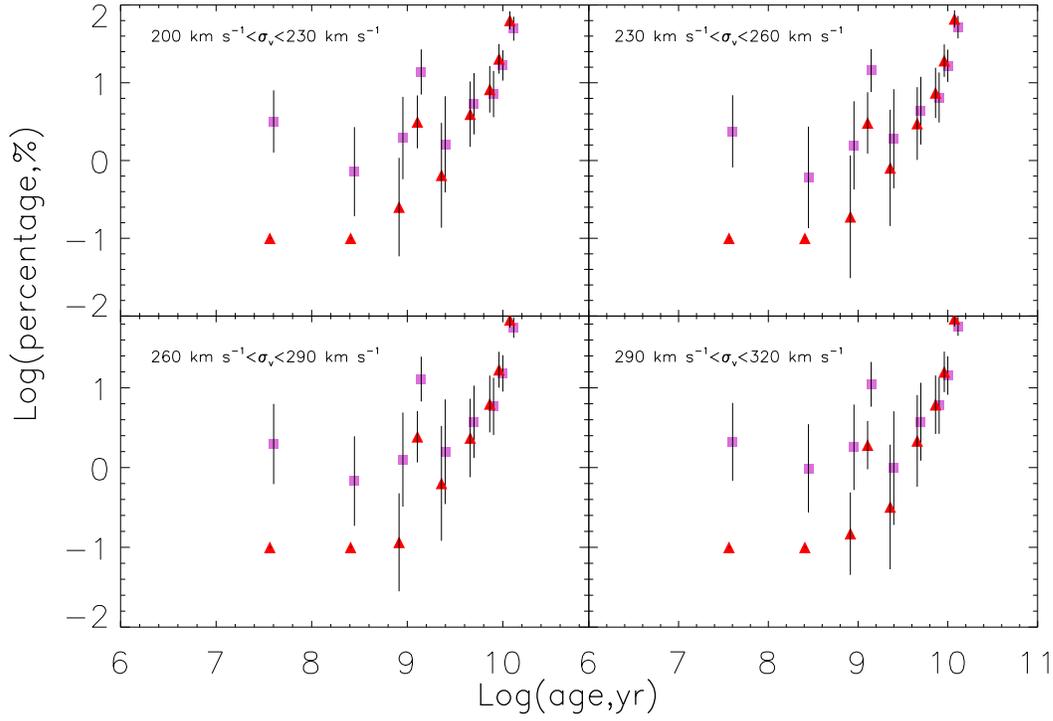}
\caption{ Star formation history of LRGs. Squares and triangle represent the light-weighted and mass-weighted percentage, respectively.}
      \label{fig:f10}
\end{figure}

\begin{figure}[!htp]
\centering
\includegraphics[width=14.0cm, angle=0] {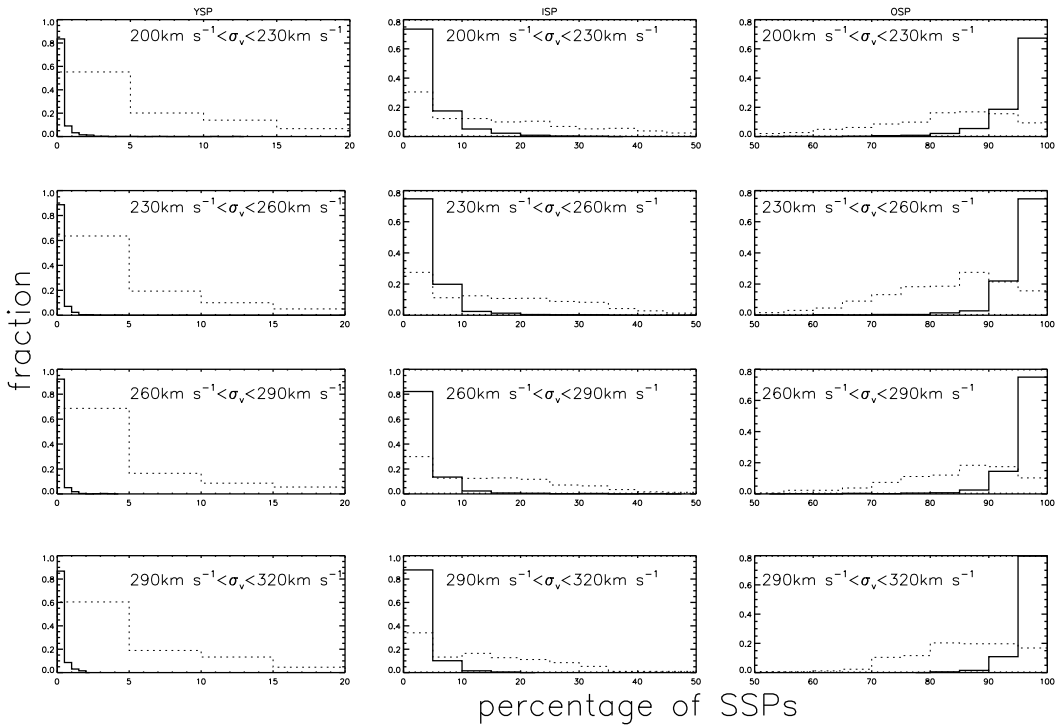}
\caption{ The distribution of YSP,ISP and OSP for LRGs. Solid line and dotted line represent the mass percentage and light percentage, respectively.}
\label{fig:f11}
\end{figure}

The star formation history of LRGs can be reconstructed according to the SSPs obtained from the spectrum fitting.  We carry on statistics on $x_j$ and $\mu_j$ from fitting for every spectrum of our sample. First of all, we sum up $x_j$ and $\mu_j$ of all resultant SSPs with different metallicities but the same age to be the contribution of this aged SSP to the flux and mass of observation at the normalization wavelength ($\lambda$=4020$\AA$). Secondly, we calculate the mean and standard deviation for the summed $x_j$ and $\mu_j$ in logarithmic coordinates. In statistics, We show percentages of SSPs with different ages for each of the 4 sub-samples in Figure ~\ref{fig:f10}, which marks the light-weighted results as symbols of squares and the mass-weighted results as symbols of triangles. For a clear comparison, we plot the mass-weighted results with an offset of 0.04 along the horizontal axis. We only designate the upper limit of fractions of very young stellar populations (4, 286 Myr) as their contribution is quite small ($\leq$ 0.1$\%$). Generally speaking, old stellar populations are the dominant component. In order to display distributions of light fractions and mass fractions of stellar populations with different ages, we divide the resultant 9 SSPs into 3 groups of young stellar populations (YSP; $\leq$ 1 Gyr), old stellar populations (OSP; $\ge$ 2.5 Gyr), and intermediate-age stellar populations (ISP; 1$\sim$ 2.5 Gyr). We calculate the light and mass fractions of YSP, ISP, and OSP for every spectrum, and show the mass fractions (solid line) and light fractions (dashed line) of YSP, ISP and OSP for each of the 4 sub-samples in Figure ~\ref{fig:f11}. From Figure ~\ref{fig:f11}, we can conclude that, for YSP, a large fraction have mass fraction less than 1$\%$ and light fraction less than 10$\%$; for ISP, most have mass fraction less than 10$\%$ and have spread distributions of light fractions; for OSP, the majority have mass fraction more than 90$\%$ and light fraction from 50$\%$ to 100$\%$. On the whole, old stellar populations formed in an early peroid and with less than 1$\%$ of mass formed in low redshift universe are the dominant components in LRGs. All these conclusions completely agree with those from previous work (\citealt{Cimatti+etal+2008};\citealt{Spinrad+etal+1997};\citealt{Thomas+etal+2010}).

\section{Summary}

We compare ULySS and STARLIGHT by using them to simultaneously study the stellar populations of a sample of quiet LRGs selected from SDSS DR7.
For those LRGs with high S/N spectra, the two packages can give generally consistent results, although ULySS may give older ages and richer metallicities. For LRGs with low S/N spectra, results obtained from the fitting by ULySS need to be checked carefully because ULySS can be easily trapped some local minimum region of parameter space. Based on the fitting results of high-S/N spectra, we further investigate the age-$\sigma$ relation and star formation history of LRGs. We find that the  majority of LRG mass formed at high-redshift, and the ages of LRGs increase with increasing velocity dispersion, which is consistent with the 'downsizing' formation picture of galaxy.

\normalem
\begin{acknowledgements}
This work was supported by the National
Natural Science Foundation of China  (11033001 and 11073024).
\end{acknowledgements}

\bibliographystyle{raa}
\bibliography{bibtex}

\end{document}